# Arduino Tool: For Interactive Artwork Installations.

Murtaza Hussain Shaikh

*Department of Computer and Information Science (IDI),*
Norwegian University of Science & Technology (NTNU), Trondheim 7448 - Norway.

**ABSTRACT**
The emergence of the digital media and computational tools has widened the doors for creativity. The cutting edge in the digital arts and role of new technologies can be explored for the possible creativity. This gives an opportunity to involve arts with technologies to make creative works. The interactive artworks are often installed in the places where multiple people can interact with the installation, which allows the art to achieve its purpose by allowing the people to observe and involve with the installation. The level of engagement of the audience depends on the various factors such as aesthetic satisfaction, how the audience constructs meaning, pleasure and enjoyment. The method to evaluate these experiences is challenging as it depends on integration between the artificial life and real life by means of human computer interaction.

This research investigates *"How Adriano fits for creative and interactive artwork installations?"* using an artwork installation in the campus of NTNU (Norwegian University of Science & Technology). The main focus of this investigation has been to get an overview on the intersection between information technology and Arts. This gives an opportunity to understand various attributes like creativity, cooperation and openness of processes influencing the creative Artworks. The artwork is combination of Adriano and other auxiliary components such as sensors, LED's and speakers. The data for analysis is mainly collected through the questionnaire to people interacting with the installation. This data collected proved that the choice of hardware and software plays a vital role in shaping the creative artworks; the investigation has also showed that there are other factors as well which will influence the creativity such as budget, selection of tools and design for the artwork.

## 1. INTRODUCTION

As the computer science field has grown to be ubiquitous part of human's life, the need for making software and hardware support to fit to the needs of the users is constantly increasing. The issue closely connected with this phenomenon is creativity. Nearly every activity is involved with the creativity. Creativity occurs when somebody produces work which is useful, generative and influential [1]. While creativity is involved with every work but it is difficult to access the creativity as creativity differs from person to person. The creative work from one person may seem to be general work from others point of view. The historical and Psychological attributes are also important to define the creativity [2]. The emergence of the digital media and a computational tool has opened the doors for creativity [3]. The cutting edge in the digital arts and role of new technologies can be explored for the possible creativity [3]. This gives an opportunity to involve arts with technologies to make creative works. There are other factors which influence the creative work, one such factor is the interactivity. The interactivity defines that artist; artwork and user have a role to play in order to give shape to the output of an interactive artwork [4]. The information technology have been influencing in shaping the creative artworks from 1960's. There is an intersection of Arts and Software interests which attracts the people from different backgrounds to work together for creative Artworks [5]. The tools available these days are very well suited to fulfill the needs of people from different background. There are many open source software and hardware tools available explicitly for creative works, one such tool is Arduino. An Arduino is an open source electronics prototyping platform configured with electronic components such as micro controllers, LEDs, sensors etc. and it can be used for creating interactive objects and environments [6].

## 2. PROBLEM DESCRIPTION

The main focus is to get an overview on the intersection between information technology and Arts. This gives an opportunity to understand various attributes like creativity, cooperation and openness of processes influencing the creative Artworks [7]. As a result of this work will mainly focus on enhancing state-of- art technology and research on the collaboration between software and arts with its implementation addressing the environment issues like water pollution, this will lay a good foundation for the future work on Role of creativity in shaping the Artworks. With the above stated goals, we propose the following research problem statement for this project;

- *To study the Arduino for creative and interactive Artwork installations.*

More concretely, the research is carried out to understand the collaboration between art and software using Arduino based artwork with water pollution as case study of research. The research is carried out with an installation which has a message about side effects of the water pollution. The primary goal of this paper is to understand the interaction between Artwork, Artist, Visitor and software / hardware phenomenon of ArTe conceptual framework [8]. As the artwork is concerned with the environment issue of water pollution, there is possibility to extending the research further to;

➢ *Understand the collaboration between the software and arts in solving environmental issues.*

➢ *Educating the people about the environmental concerns and at the same time encouraging them to be creative.*

## 2.1 Research Methodology
The purpose of the research was to get a deeper understanding of the various elements involved in the creative and interactive artworks; the research method used was explorative, as the explorative study is helpful for the researcher to understand the research problem [52]. This method gives an opportunity to get better understanding of one has not aware on and also gives an opportunity to researchers to base their work on previous existing works [9]. The research was conducted basically by using two types of research methods: the first research method was to explore the literature to get an understanding of the current situation of the interactive artworks and usefulness of the tools such as Arduino in creating interactive artworks. The literature study being explorative, rather than systematic provides a platform to discover unknown and deeper aspects of the artwork installations. Secondly, the research involved the installation of the artwork and thereby questionnaire was performed to get find the interactivity of the Artwork.

## 2.2 Research Scope
The scope of this research has been narrowed by the problem definition to understand the usefulness of the Arduino in making the creative artworks and factors influencing the produced artwork. The concept behind the creativity is wide; it depends on the field and nature of the project chosen. In this research we are addressing the creativity involved with the artwork being developed. Also there is interesting aspect with the installation; the people interacting with the installation can also use their creativity to use different artifacts like paper, pen etc to check the output from the Artwork. The research is carried out by using an Art installation in a public place, which comprises of Arduino and multimedia technologies. The multimedia technologies have also changed the process involved with art and the usage of music, video and displays in making an appeal to the user [10]. However, this research does not cover the role of multimedia technologies in shaping the interactive installations. The results from this research can be used for analyzing some of the main issues discussed in Software Arts such as educational and social issues of art and software [10]. The data collected from the project will enable us to fully understand the hardware and software role in shaping the creative artworks and understand the various aspects influencing the interactive Artwork installations. Furthermore, the research can be conducted in the future by using the data collected in this project. The data collected in this research can be useful for analyzing the involvement of Artists in the software development and awareness of people towards environment pollution as the project involves artwork installation related to the water pollution and participant's intuition while working with installation.

## 3. LITERATURE REVIEW ON ARTS
To perform research through the artwork installations, it is essential to understand the various elements involved in the artwork installations. As we see today there has been growth of arts from earlier times to current time. The role of technology is more prevalent these days than the earlier times. The technology has removed the complications involved with the artworks and making it feasible to deploy artworks for the people in different fields.

## 3.1 Historical Considerations
The term ART is often confusing; the meaning of the word has been changing from time to time. The work which is considered as an artwork may not have considered as a piece of artwork when it was first designed or the person who have made the work may not have considered it as an Artwork. The notion of "art" and "artist" are relatively modern terms [11]. The Art does not have any concrete definition, according to the Britannica online, Art is "*the use of skill and imagination in the creation of aesthetic objects, environments, or experiences that can be shared with others*" [11]. Therefore, the art can be considered as the *"work which produces some aesthetic sense"* [12]. At Renaissance phase, the Art emerged as collective term comprising Painting, Sculpture and Architecture [11]. This has further expanded to the other areas such as Music and Poetry, which became part of the fine arts in the 18th century. So, these five fields formed as a fundamental unit of fine arts separating them from the decorative "arts" and "crafts", such as pottery, weaving, and furniture making [11]. This arise few questions like how art can be distinguished from crafts. What is the difference between artist and craftsmen? In the ancient days, the term we consider, Art today was an activity governed by the rules [13]. The activities such as Painting, Sculpture and Weaving were considered as human activities rather than crafts [13]. In the latter part of the 19$^{th}$ century and the beginning of the 20$^{th}$ century, the notion of truth to one's material emerged. It is called as *"Art for Art's Sake"* [14]. In the early 20th century, the notions about the art and Artists were penalized and the Art was redefined as *"Anything which is produced by an artist is an Art"* [11]. This has made the Art as a broad field, which includes different elements involved in it such as Artists, Artworks, Technologies and people involved with the artworks, the perception of the participants or observers and so on. This has made the communication as the key aspect to distinguish between a work and Artwork. The work which has an appeal to the observers or which communicates a message to the observers interacting with it can be called as an Artwork [15].

*3.2 Current Paradigms of Arts*
The form of art usually differs from the type of Artwork such as painting, drawing, sculpture, architecture, electronic media such as computer and digital graphics, ceramics, Visual Design, Graphic Design, photography and so on [16].

*3.2.1 Painting*
Painting is a graphic art which is made by the applying paints to a surface using the artistic composition [17]. The painting is in existent from the ages. Starting from the cave paintings of Lascaux to some great masterpieces of Leonardo Da Vinci, the arts have played a historical and aesthetic role in all the ages [18]. In figure3.2.1 one of the famous paintings named Monalisa in the renaissance phase is illustrated. The paintings developed into many different types such as historical, allegorical, religious, portrait and so on over the years [18].

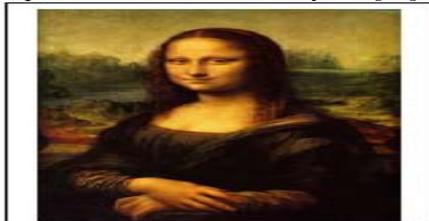

*Figure 3.2.1 : Monalisa, masterpiece from the painter da vinci in the renaissance period [19]*

*3.2.2 Drawing*
Drawing is an Art or technique of producing images on a surface, usually paper, by means of marks in graphite, ink, chalk, charcoal, or crayon [20]. The drawing has always been part of the art, it is considered as a language for expressing the nature and creativity [21].

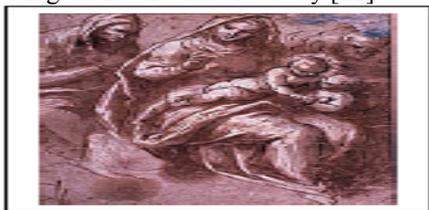

*Figure 3.2.1: Madonna & Child, c.1580, Agostino Carracci work [22]*

*3.2.3 Photography*
The photography is a process of producing visual images of the objects on photosensitive surfaces, which can captures the image [23]. The process of photography is relatively new to world of art [24]. The arts in renaissance period grown to large extend based on the criteria to capture the nature, realism and the emergence of the photographic devices, the extent to capture the images has been increased by leaps and bounds. The details which cannot be portrayed by an artist can be possibly captured with a photograph.

*3.2.4 Sculpture*
The art or practice of making figures and designs using tools on stones, casting on metals or modeling using clay [25]. The sculpture has been the primary source of expression from thousands of years. The sculpture is often used to represent the societal concerns, such as religion, politics, and morality [26].

*3.2.5 Digital Art*
Digital art is a form of contemporary art, making distinctive works using the computer technology [29]. This art emerged in the 1970s but came into the limelight after growth in the technology like computers, software's, video devices, sound mixers and digital cameras. The process of digital art has been associated with various names computer art and multimedia art but the digital art is placed itself under the broad category of Media Arts [30]. There are different forms of digital arts such as Algorithmic art, Art game, Art software, Computer art, Computer art scene, Computer generated music, Computer graphics, Computer music, Cyber arts, Demo scene, Digital illustration, Digital imaging, Digital morphogenesis, Digital painting, Digital photography, Digital poetry, Dynamic Painting, Electronic art, Electronic music, Evolutionary art, Movie special effects, Fractal art, Generative art, Immersion (virtual reality), Interactive film, Machinima, Motion graphics, Multimedia, Music visualization, New Media Art, New Media Photo manipulation, Pixel art, Software art, Systems art, Traditional art, Video art, Video game art, Video game design, Video poetry and Virtual arts [30].

## 4. ARTIST CREATIVITY ISSUES

Having seen the different forms of arts which form the basis for creating Artworks today, it is time to focus on the creativity of the Artist itself. We will not focus on the creativity among the artists to do the artwork instead we will focus on the hindrances which artists might encounter while creating their artworks. There is however no secret that some of the works related to the painting, photography etc has been under criticism during the past centuries. It is very important to understand that creativity is often involved with the Artworks.

## 4.1 Artworks and Creativity

The artworks produced by an artist can have different purposes; one of them can be interactivity i.e. to engage the audience. The interactive artworks are often installed in the places where multiple people can interact with the installation, which allows the art to achieve its purpose by allowing the people to observe and involve with the installation [35]. The level of engagement of the audience depends on the various factors such as aesthetic satisfaction, how the audience constructs meaning, pleasure and enjoyment [36]. The method to evaluate these experiences is challenging as it depends on integration between the artificial life and real life by means of human computer interaction [37]. The artworks can be creative, which involves the participants to work with the installation to produce some new functionality which can be further developed or which creator of the artwork has not thought of. This creates the necessity to understand the creativity thought process development and the life cycle of the creativity process in order to design the artwork having creative edge. The following sections present the development cycle of creativity, and then we will present the constraints of creativity and various issues relevant with the creative artwork installations.

## 4.2 Creativity Development Cycle

Creativity is often embedded in every activity we perform, Creativity can be characterized as a process towards achieving an outcome recognized as innovative [33].

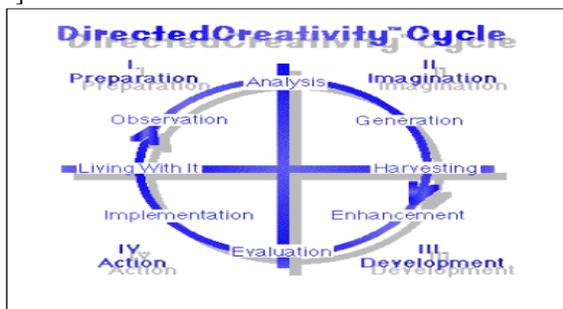

*Figure 4.2: Directed Creativity Cycle [38]*

The creativity and innovation can be represented in the form of different stages of development as illustrated in the figure 6. The stages involved in the creativity development as illustrated as Directed Creativity Cycle [38] is a synthesis model of creative thinking. The four phases involved in this creative thinking cycle are Preparation, Imagination, Development, and Action. The following is the brief description of these phases;

### 4.2.1 Preparation

Creative thinking begins with careful observation of the world coupled with thoughtful analysis of how things work and fail [38]. The preparation phase starts with the observation i.e. the inspiration and thought to produce a creative art work, it happens by observing our experience and creativity is triggered [39].

### 4.2.2 Imagination

This is the phase of inspiration and creative action [39]. The major processes involved in this phase are Time in, Generation and Harvesting. The "Time in" process means the time of silent reflection [38], which includes the process where we start to train our mind for the new creative work, for instance, we make our mind like a blank page which can be explored for any new works [39]. The "Generation" process is the stage where the creativity is tried to infuse into the process of work. This stage allows the user (i.e. person producing the creative work) to brainstorms or uses mind maps to get overall view of the creative work to be established [39]. The "Harvesting" is a process where the creative thoughts are made available for the use. In other words ready to be harvested into proper plans for actual implementation [39].

### 4.2.3 Development

This is the process where the ideas formulated from the previous phase are put into actual development. There are two major processes involved in this phase namely Enhancement and Evaluation [39]. The "Enhancement" means the process of editing, which allows the user to modify the thoughts to fit into the actual development [39]. The "Evaluation" is the process where the creator reasons himself whether he/ she is achieving what is desired [39].

### 4.2.4 Action

This is the last phase of the cycle which focuses on the actual implementation of the creative thoughts and sharing the work produced with others [39]. The major process involved in this phase is the "Implementation" which means the task of giving lie to the idea [39]. This last phase makes the creator to get new thoughts from the evaluation done by the other persons or the interaction mechanism involved between the people and work produced. This phase leads seamlessly to the beginning of another creativity cycle [39]. We have seen the various processes involved in the development of the creative art works. However, there are other aspects which will affect influence the creative work produced by an Artist (i.e. creator of the artwork). In the next section we will present these aspects which influence the artworks produced by an artist.

## 4.3 Creativity Constraints

The creativity is often bound by the constraints, the true innovation happens when the constraints and challenges are surpassed [34]. The following are the various constraints in producing creative works. The creativity is often bound by the constraints, the true innovation happens when the constraints and challenges are surpassed [34]. The following are the various constraints in producing creative works.

**CHANGING PERSPECTIVES:** The perspectives of the people varies as the creative work from one person may not be creative to other person, it can be general work from his/her perspective. The main challenge is to overcome this barrier and produce the work with outstanding creativity. The outstanding creative work is any work which has stood test of time and recognized by people beyond the specialist community [40].

**EXPERTISE IN TOOL:** The other challenge faced by the artists is the need to become expert with the tool in order to produce creative work. The tools such as image editing, modeling tools requires the user to have good understanding of the various functionalities of the tools and Programming knowledge in order to increase the horizons of their creativity.

**SELECTION OF TOOLS (HARDWARE/ SOFTWARE):** The choice of the tool depends mainly on the creativity of the artist. The creativity often depends on the selection of the tools for making the creative artwork. However, there can be some factors which will prevent the artist using the tools such as not meeting the full requirements of the project or lack of tool expertise.

## 4.4 Meeting User Expectations

The creativity is also dependent upon the expectations of the users i.e. Artists. The Artists often look for the processes or factors which make the project (or Artwork) feasible. The following are some of the expectations which will make the Artist to produce creative work.

**LOW COST:** The cost plays a major role in designing and planning the artwork. The Artist imagination and creativity are often hampered by the budget allocated for producing the artwork. In order to make the artist feel comfortable with the artwork project is to make the project composed of the things which are viable and easy to use. One such method is to use the open source, which will drastically reduce the cost for the artwork project and also provides the flexibility for the user to change or modify according to the needs of the artist as per GNU GPL license provided by the open source.

**COLLABORATION AMONG MULTIPLE FIELDS:** The growth in the technologies has made possible to produce artworks involving knowledge from different fields. For instance, the interactive installations developed by the artists such as Samir M.kadmi artwork- Sonic Onyx [41], Øyvind Brandsegg artwork Flyndre [42] and so on. These installations are built using knowledge from many fields such as mechanical, electronic, programming etc. This illustrates that the scope of the artworks are pervasive and it is expanding to various fields. In order to make these installations possible the collaboration between different people from different fields is important. It is fruitful to open research centers such as COSTART, a research center based on multi-disciplinary foundations such as Art, Design, Science and Engineering to improve the human–computer interaction and facilitate artwork installations using collaboration between people from different fields [3].

## 5. ARDUNIO TOOL

Arduino is an open-source electronics prototyping platform having both hardware and software components [6]. The main intention for designing the Arduino platform is to suit the needs of the artists, designers, hobbyists, and anyone interested in creating interactive objects or environments [6]. The Arduino board consists of a micro controller which enables to receive inputs from the sensors and thereby can be used to drive motors, LEDs, sensors and other components [45]. Microcontrollers are small computational devices embedded on integrated circuit containing processor, memory and other peripheral I/O pins [46]. The figure below illustrates one of the Arduino Diecimila, one of the microcontroller board based on the ATmega168 [6].

The microcontrollers have existed from decades, However Arduino microcontroller is explicitly designed for artists and designers [45]. The main advantage for the artists and the designers is to execute without knowing the internal functioning of either hardware or software of the Arduino [45]. The Arduino was aimed to be used by non technical users. The Arduino was made apart from other microcontrollers by providing four major features they are inexpensive, providing an IDE for code development, programming via USB and community support for Arduino [6].

## 5.1 Purpose of Artwork Installation

In recent years, artists are exploring possibilities to extend arts with the computation [31]. The role of information technology is increasing in creating the artworks. The creativity also plays an important role in choosing the right technologies for the artwork.

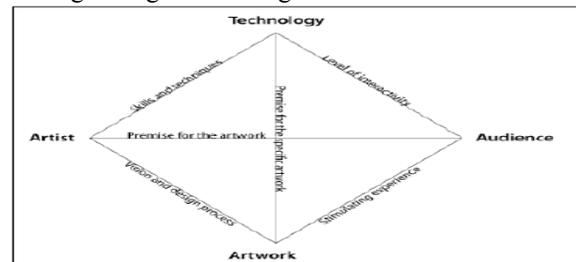

*Figure 5.1: ArTe conceptual framework*

There are various aspects associated with the interactive artworks and relations among those elements are important to analyze the interactivity of the artwork. The aspects are Artists, Artwork, Technologies and participants. The relation between these factors can be studied using the ArTe conceptual framework as illustrated in the figure 5.1 [32]. The focus on sharing, cooperation and openness is central in the artwork project, particularly with regards to technology.

### 5.2 System Description
The following are the design aspects of the created artwork and its working procedure.

**SYSTEM ARCHITECTURE:** The system architecture is illustrated in the figure below, the installation includes a blue strip of paper depicting the river and also Arduino micro controller is connected to installation along with audio speakers, LEDs and Flash memory. The LEDs are placed underneath the river (i.e. blue strip of paper) connected in such a way that it will change the color depending upon the signals or input received from the participant (or Artist) working with the project. The participant is provided with a scenario of having recycling containers and some toxic containers (Note: the containers refer to the vessels holding the garbage) near the water body. The participants are provided with some artifacts readily available at installation area like paper, polythene covers, oil or any liquid. When the participant puts those objects in the containers , the IR sensors which are present in the containers will detect the objects being placed and then the LEDs will change their color and also the color of water (i.e. in our case it is blue strip) flowing in the artwork (as LEDs are placed underneath the water). This project also provides scope for the participant to have creativity edge, by allowing participant to use other objects to place inside the container. The participant can also sense the music being played upon adding pollutant to the water. The audio output will differ from pleasant music like birds singing or any melody depicting happy moments to sad songs depending upon the degree of pollution added to the water. This feature will help the participant and nearby visitors to hear and imagine with respect to the side effects of pollution.

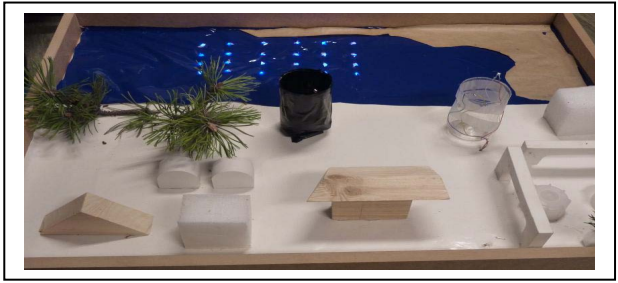

*Figure 5.2 a: The Architecture*

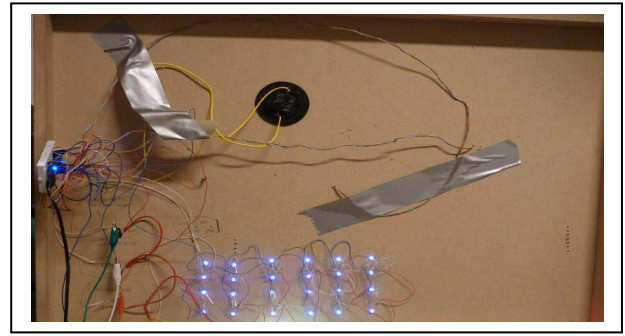

*Figure 5.2 b: The Installations*

**HARDWARE:** The artwork is mainly built on the Arduino microcontroller platform and thereby can be divided into majorly three functionalities firstly, the LEDs and second one is the sensors and third is the Audio Speakers. The Sensors are used to track the objects being thrown into the containers and upon detecting the changes in the infra red sensor values the color of light in LEDs changes. Also the user will be able to hear the audio from one of the speakers in response to the sensor values; the sensors are placed in the containers to collect the garbage, where one belongs to the environmental friendly container which will not pollute the water and other being the toxic container, which may pollute the water. The user throws any object into the environment friendly container a good melody with birds singing is played otherwise the audio changes to thunders and storms, to illustrate that something is wrong and is not good for the environment. The interactive part of the artwork is when user puts anything in the container there the artwork will 1) The sensor detects the user input and container and 2) changes the color of the LEDs and 3) The audio is played in response to the user and container where the objects are put into.

| Part | Specifications / Tools for Configurations |
|---|---|
| Micro Controller | Arduino Nano, Mini-B USB cable |
| Lighting System | LEDs COM-09264 ( Triple Output LED RGB - Diffused - 1pcs) |
| Sensor | IR sensors SEN-00241 (Infrared Emitters and Detectors) |
| Audio System | Audio speakers COM-09151, Flash memory connector DEV-09534 ( Audio-Sound Module - SOMO-14D) |
| Other | Bread board, Connectors / wires |

*Table 5.2: Specification for tools*

### 5.3 Artwork assessment using ArTe Framework
The creativity support tool *(such as Arduino)* can be truly tested by considering the impact level of tool on the creative thought process. The assessment has to be based upon various criteria`s such as efficiency of the tool, cognitive support and other subjective assessments [56], which are governed by the various phenomenon such as Artist, Artwork, Visitor and Software, The ArTe conceptual model is based upon these phenomenon. Therefore the ArTe conceptual model can be applied to Arduino, a creative support tool to understand the collaboration between the various phenomena associated with tools for achieving the creative works. The following table illustrates the relationships that are associated with Artwork as per the ArTe framework.

| Tools / Hardware | Artist | Artwork | Visitor | Software |
|---|---|---|---|---|
| *Arduino Nano*, Mini-B USB cable, LEDs, IR sensors, Audio speakers, Flash memory connector, Audio-Sound Module, Bread board, Connectors / wires etc. | Creator of the artwork, Participants interacting with Artwork | Completed Artwork (as illustrated in fig. above) | Persons present in the installation area, outside persons visiting the installed area | Processing / C programming language, Arduino IDE |

*Table 5.3: Relationships with ArTe framework.*

## 6. QUESTIONNAIRE DESIGN

The questionnaire is effective way to collect the data from many people for research [52]. This method does not often give a chance to go back and ask the questions [52], so we have designed the questions properly to increase the response rate and probability to get more data for the analysis. The questionnaires can be classified mainly into four parts namely, Factual data (basic information about the participants), Interactivity of the Artwork, Hardware and Software fulfilling the requirements of the project, Limitations and scope for improvement.

***Factual Data:*** These questions were mainly to know the background information of the person such as education level, department of study, age group and gender. This information is essential in order to be aware of the sample of people contributed to the data and their background related to the Arts or the engineering.

***Interactivity and Creativity of Artwork:*** These questions were designed to check whether the users were able to interact with the Artwork or not. Also there were few questions related to the creativity such as whether they were able to use the nearby objects with the artwork in order to check the output. The questions were basically closed, where the users are given an option to select the choices from pre defined choices [52]. In the latter question of this part, user is given a choice to say YES / NO to get the feedback on the creativeness of the artwork.

***Elements of Artwork:*** This part of questionnaire is crucial part of the artwork, as we wanted to find out the role of the hardware/ Software and other elements such as wooden frame or any part involved in the artwork. The questions were mainly degree of agreement questions, as known as "Likert Scale" [52] and some were scaling based questions such as very interesting to none interesting, in order to truly find the purpose fulfilled by various elements in shaping the creative artwork.

***Limitations and Scope of Artwork:*** This part of the questionnaire was to find the limitations of the artwork and thereby find out the improvements and further development in order to make it suitable for wide audiences. The majority of the questions were quantity questions such as are there additional features that you would want incorporated into this Artwork? And some of the questions were also based on the closed type [52], such as what are the features required to make it more creative?

### 6.1 Participants

The participants required for collecting the sampling or questionnaires with the students in the university seemed as the most probable method, we have chosen the students at researcher's workplace. This process may not be enough to test the artwork, but it was selected for various reasons. Primarily it fits our analysis where students being from the computer science department, they are somewhat aware of the technologies and tools. Also due to the time constraint and majority of the students from other fields are taking their exams, so we have chosen to perform the research using the data collected from these chosen set of people.

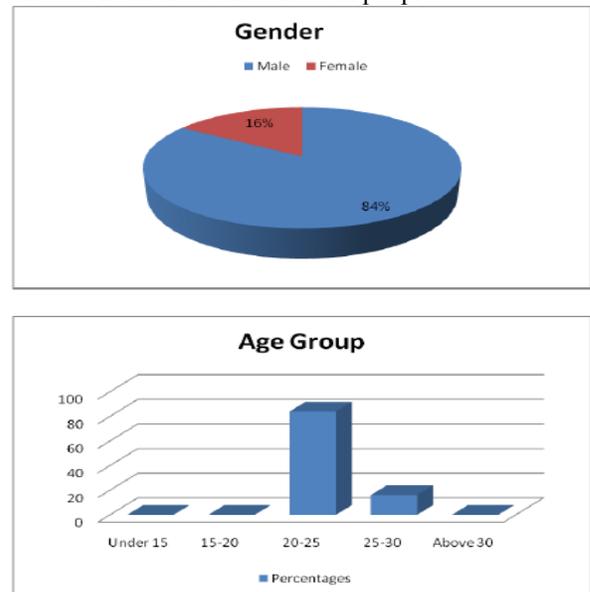

*Figure 6.1: Age group and Gender*

In total 7 from the participated 9 students filled the questionnaires, where 5 persons filled completely and 2 persons filled partially and thus yielding a response rate of nearly 80%. Out of the incomplete 2 responses, one was rejected as they have not answered the essential part of the research. This gave the final sample of 6 people, where 84% were the male and 16 percent were female. 84% of the students (i.e. Participants) were in the age group of 20-25 and only 16 percentages were in the age group of 25-30.

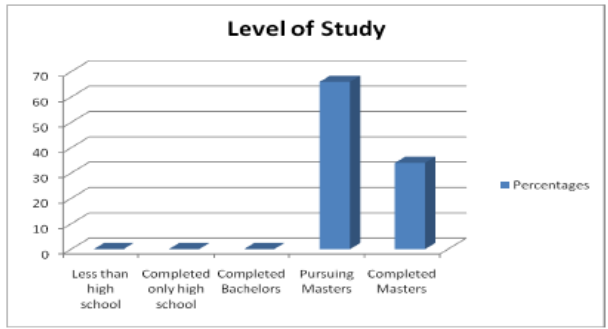

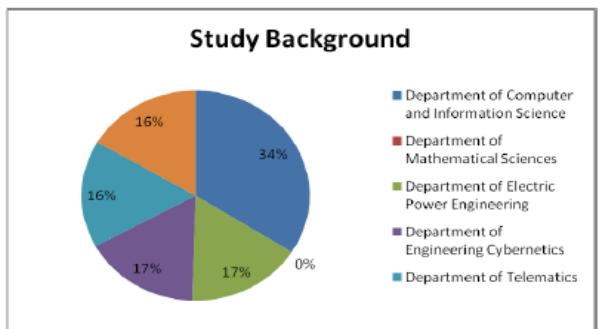

*Figure 6.1.1: Study Background and level of study of Participants*

The background of the students were highly educated and majority of the students had already completed the degree either bachelor or masters. All the students were associated with computer science department taking various programs of study at the university.

### 6.2 Results

As can be seen in the figure 6.2 the responses from the participants illustrate that the majority of the people are aware of the theme f the project as related to the recycling process and the theme as the water pollution. The percentages for the people thinking as recycling process were 50 where as 34 felt it as water pollution, while 16 percent of them felt that it has both the shades of recycling process and the water pollution.

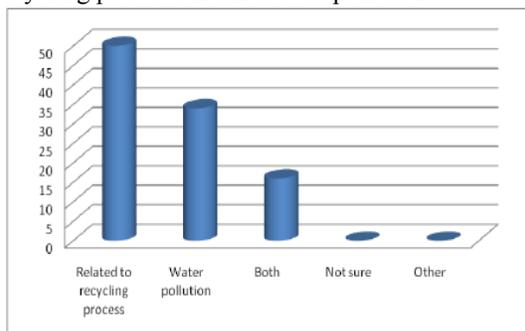

*Figure 6.2 : Responses to "What is the Purpose of the Project"*

Not surprisingly, the responses from the people were positive in terms of the interactivity as can be seen in the figure 6.2 , 34% of the respondents felt that the interactive and 50% as partially interactive, where as 16% as non interactive.

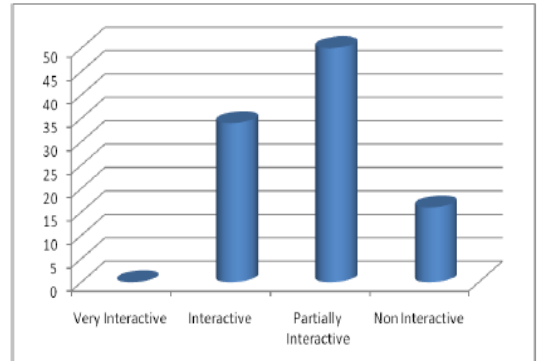

*Figure 6.2.1 : Responses to "Interactivity of the Artwork?"*

This is an expected result as the sensors were placed inside the container and people were trying to use nearby objects to place in the containers. The responses as can be seen in figure 6.2.1 prove this fact; nearly 66% of the respondents were able to use the nearby objects to put into the container.

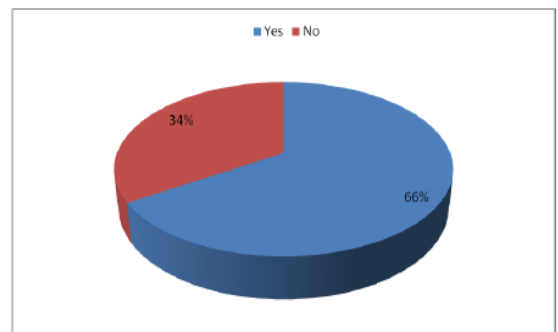

*Figure 6.2.2 : Responses to "Were you able to use the nearby objects to place in the container, Is this creative?"*

***Elements of Artwork:*** The purpose of this part of the questionnaire is to access the various parts related to the artwork. As can be seen in the figure 6.2.3 the opinions are quite in agreement to what we have discussed in the preliminary studies. The reason for these positive responses can be because of the impression made to the user as "correct" answer [55]. In the results, we can see the pattern in the responses and the understanding of the users. In the responses for the layout and design of the artwork, the responses were mostly ambiguous as nearly 33.3 percent for somewhat important, least important and 33.3 LED`s and unsure about the layout of the artwork. In case of Sensors, people shown some positive signs as majority of the respondents 34% felt that it was either somewhat important or important and 16% felt that it was most important. However, 16 percent have responded for least important. In the case of speaker, the respondents felt that it is important as 34 percent for important and 50% for somewhat important, where as 16 percent for least important. For the LED`s, nearly half of the people felt that important.

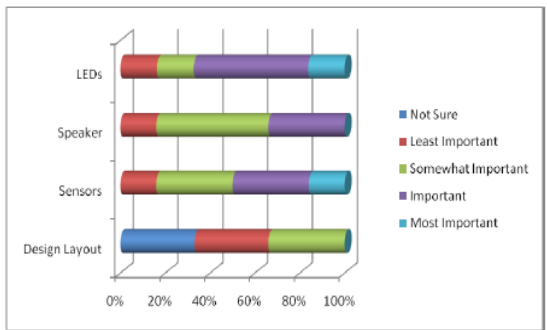

*Figure 6.2.3 : Responses to "How important are the following features of the Artwork?"*

In order to completely understand the drawbacks of the created artwork and the importance of the artwork, the question was asked on the interesting aspects of the artwork, thereby and the responses were for the LED`s, 34% felt that it is either uninteresting or neutral and 16 percent felt that it is interesting and also not sure of. In the case of the audio output from the artwork, 16% for the interesting and 34% of the response for both interesting and neutral. For the Containers placed in the artwork, the majority of the responses were neutral as 50% felt neutral and 16 percent for uninteresting and also unsure about the containers. For the design of the artwork, the responses were mostly towards the unclear and uninteresting as 34% were unsure and 16 percent were uninteresting, 34% were also neutral about the design.

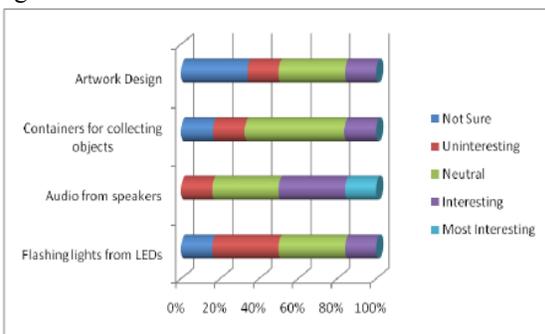

*Figure 6.2.4: Responses of "Would the following features make you more or less interested in Artwork?"*

***Limitations and Scope of Artwork:*** In order to test the scope of the project and future implementations, the responses were asked about the features which can be added to make it more creative and the factors effecting the development of these artwork projects and its implementation to the needs of the users. As can be seen in the figure 6.2.5 the responses were mixed as it can be seen nearly 34 percent were neutral about the artwork to their needs and remaining others were spread from very likely to unlikely.

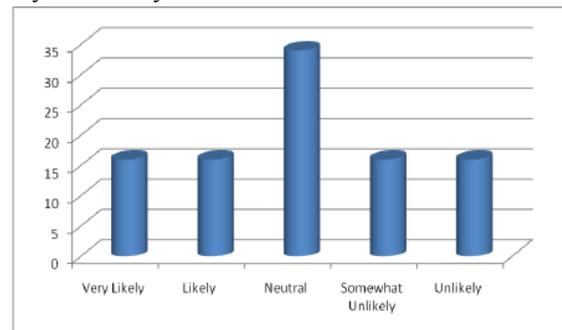

*Figure 6.2.5: Responses to "Are you interested in developing these kinds of artworks for your needs?"*

The responses for the factors for developing the artwork can be seen in the figure 6.2.6 the responses for the reasonably priced, 34% were stating that they are in agreement either strongly or somewhat. 16 percent for both neutral and somewhat disagree. In the case of the Collaboration with people from other fields to complete the artwork, 16% were strongly agreeing and 50% somewhat agree. Whereas, 34% of them were neutral. For the selection of tools for design of the artwork, the responses were mostly uniform as 50% were neutral and remaining were either agreeing partially or disagreeing with it.

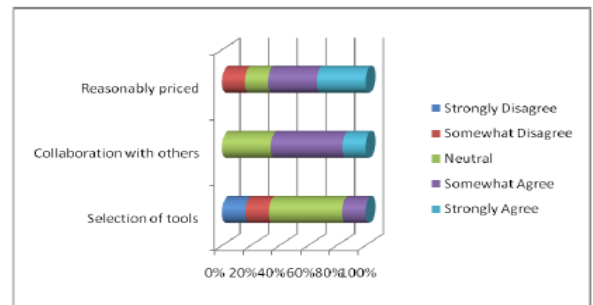

*Figure 6.2.6: Responses for "Developing the Artwork by themselves"*

The responses for the needs of the creative artworks as can be seen in figure 6.2.7 are almost same as 34 percent felt that it should meet the user expectations, user friendliness and better selection of tools.

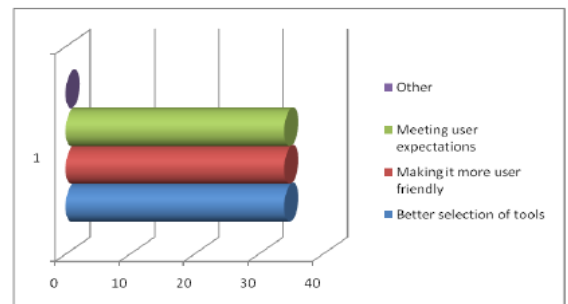

*Figure 6.2.7: Responses to "What are the features required to make it more creative?"*

## 7. VALIDITY AND LIMITATIONS

Our findings are purely indications of users (i.e. students in the computer science department IDI) experiences and their interaction with the artwork, and should not be understood as evidence. Possible sources of bias in the questionnaire may result due to the age, nationality and education. In addition, the fact that respondents in the questionnaire are related to the researcher as "friends" may have influenced the responses as well, for instance, answering the questions as may be "correct" instead of giving an honest answer. However, the responses from one person are disclosed from another in order to avoid this. If the artwork is installed in different public places and gathering responses from wide probability sample, which could have provided us more in depth analysis of various questions mentioned in the questionnaire. However, with respect to the purpose of the project, we feel that the questionnaires have served their purpose to get a response on the different aspects as found during preliminary studies of the project. To conclude, we feel that the results are sufficient to perform the analysis and can provide a good foundation for further research in this field.

## 6. CONCLUSION

The purpose of the Artwork installation is to enhance the knowledge of the participant's over the environment issues concerning water pollution, thus leading to increase the awareness over this water pollution problem, it seems that the participants are already aware on the water pollution and its causes to the environment. It is therefore can be seen that almost 50 percentage of the respondents in the questionnaire were able to predict that water pollution is usually effected by various factors such as garbage dumping and industrial wastage, thus understanding the role of the containers in the artwork. When compared to this 34 percent felt that the artwork is interactive and 50 percentage of the participants found that the artwork is partially interactive as there were no proper guidelines to operate the artwork, it confirms that the artwork is able to convey the message of making people aware of the water pollution issues and also artwork was interactive enough to engage the participants, thereby the artwork played as a source to provide the scenario concerning to the environment. The reasons for users not being interactive with the artwork can be related to the various factors governing the interactivity such as improper choice of the tools or technologies used lack of involvement from the user towards the artwork and so on. Although we have so many different factors influencing the artwork, majority of people felt either partially interactive or interactive and also filling the field such as additional comments in the questionnaire, participants answered that the artwork is interactive and can be improved for making it easy to understand and operate , also they answered that they were delighted by the fact that the artwork is able to communicate a message upon sensing the things being dropped into one of the container provided in the artwork. Also they were impressed by the sound when the music pattern changes upon the small things dropped into the different containers. However, the participants suggested it could have been better if it was connected directly to sensors rather than operating it manually.

Another interesting aspect of the artwork we presented was to find the creativity of the artwork, the scope of creativity provided by the artwork as either way of using the things nearby the artwork such as papers, pens or the way the participants able to imagine while working with the artwork. These are the essential factors for a creative artwork while simultaneously making the participants aware of the intention of artwork i.e. awareness on the water pollution. The observation performed illustrated that participants were able to use their creativity and thereby involve the small objects such as pen with the artwork in order to test the artwork. We have presented various factors which can influence the artwork, one such being the choice of the tools being used for the artwork. We have chosen the Arduino as a platform for the hardware required for the artwork. The role of Arduino was critical as it should support various other components such as Led's, sensors and speaker, which play a vital role in completing the artwork and its purpose. The Arduino tool fulfilled these primary aspects of providing a means of message related to water pollution and at the same time it helped in producing an artwork which is interactive. It seems that there are other factors concerning the Arduino such as technical aspects, one should possess the knowledge to use the hardware and way of connecting ports of different components such as ground, VCC and so on. The tool also provides the good interface to program, However there are other factors which has to be considered here such as the knowledge of programming, the Arduino supports the processing programming language so the artwork creator should have expertise in programming language in order to effectively utilize the programming and its functions to the needs. It has been found during development that the artwork needs knowledge in electronics in order to connect to various pins of Arduino and also good programming knowledge in order to be able to control the components and utilizing the outputs from one component to another. These aspects related to the hardware and programming of Arduino tool can be accessed from the persons, who have already used it. The data collected through this research can also be used for finding the role of Aesthetics and technologies with arts as there are different components and technologies contributing to the produced artwork.

## ACKNOWLEDGEMENT

I would like to express my gratitude to the department of Computer and Information Sciences (IDI), Norwegian University of Science & Technology (NTNU) for all the help in writing this paper. I would especially like to thank my beloved Mother Mrs.Naseem Ahsan Shaikh who always pray and helped me to improve my scientific /academic career.